\def\year{2021}\relax
\documentclass[letterpaper]{article} % DO NOT CHANGE THIS
\usepackage{aaai21}  % DO NOT CHANGE THIS
\usepackage{times}  % DO NOT CHANGE THIS
\usepackage{helvet} % DO NOT CHANGE THIS
\usepackage{courier}  % DO NOT CHANGE THIS
\usepackage[hyphens]{url}  % DO NOT CHANGE THIS
\usepackage{graphicx} % DO NOT CHANGE THIS
\usepackage{natbib}  % DO NOT CHANGE THIS AND DO NOT ADD ANY OPTIONS TO IT
\usepackage{caption} % DO NOT CHANGE THIS AND DO NOT ADD ANY OPTIONS TO IT
\usepackage{xcolor}
\usepackage{paralist}
\usepackage{float}
\usepackage{hyperref}
\usepackage{multirow}
\usepackage{subfiles} % Best loaded last in the preamble
\title{DISMISS: Database of Indian Social Media Influencers on Twitter}
\author{
    Arshia Arya,\textsuperscript{\rm 1} Soham De,\textsuperscript{\rm 2} Dibyendu Mishra,\textsuperscript{\rm 1} Gazal Shekhawat,\textsuperscript{\rm 1} \thanks{Arshia, Soham, Dibyendu and Gazal have equal contributions} Ankur Sharma,\textsuperscript{\rm 3}\\
    Anmol Panda,\textsuperscript{\rm 1} Faisal Lalani,\textsuperscript{\rm 1} Parantak Singh,\textsuperscript{\rm 1} Ramaravind Kommiya Mothilal,\textsuperscript{\rm 1}\\
    Rynaa Grover,\textsuperscript{\rm 1} Sachita Nishal,\textsuperscript{\rm 1} Saloni Dash ,\textsuperscript{\rm 1} Shehla Shora,\textsuperscript{\rm 1} Syeda Zainab Akbar,\textsuperscript{\rm 1} Joyojeet Pal \textsuperscript{\rm 3,}\textsuperscript{\rm 1}
    \\
}
\affiliations{
    \textsuperscript{\rm 1}Microsoft Research\\
    Bangalore, India\\
    \textsuperscript{\rm 2}Ashoka University\\
    Sonepat, India\\
    \textsuperscript{\rm 3}University of Michigan\\
    Ann Arbor, Michigan\\
}
\begin{document}

\maketitle

\begin{abstract}
Databases of highly networked individuals have been indispensable in studying narratives and influence on social media. To support studies on Twitter in India, we present a systematically categorised database of accounts of influence on Twitter in India, identified and annotated through an iterative process of friends, networks, and self-described profile information, verified manually. We built an initial set of accounts based on the friend network of a seed set of accounts based on real-world renown in various fields, and then snowballed ``friends of friends" multiple times, and rank ordered individuals based on the number of in-group connections, and overall followers. We then manually classified identified accounts under the categories of entertainment, sports, business, government, institutions, journalism, civil society accounts that have independent standing outside of social media, as well as a category of ``digital first" referring to accounts that derive their primary influence from online activity. Overall, we annotated 11580 unique accounts across all categories. The database is useful studying various questions related to the role of influencers in polarisation, misinformation, extreme speech, political discourse etc.
%  Databases of highly networked individuals have been indispensable in studying narratives and influence on social media. In our bid to support studies on Twitter India, we present a systematically categorised database of accounts of influence on Twitter in India, identified and annotated through an iterative process of studying networks of information flows online. We built an initial set of accounts based on the friend network of a seed set of accounts based on real-world renown in various fields, and then snowballed 'friends of friends' multiple times, and rank ordered individuals based on the number of in-group connections, and overall followers. We then manually classified identified accounts under the categories of entertainment, sports, business, government, institutions, journalism, civil society accounts that have independent standing outside of social media, as well as a category of 'digital first' referring to accounts that derive their primary influence from online activity. Overall, we annotated 11580 unique accounts across all categories.
\end{abstract}

\section{Introduction}

In the last two decades of growing  social media use, a large number of functions in the public sphere are either driven by, or entirely conducted through online communication. Politics, journalism, brand outreach are among a small number of domain spaces of communications that now have a large online component. Influencers, often individuals or accounts who command a large following online and wield influence either directly or through their ability to get second-order engagement in their extended networks. These influencers play a key role in building or propagating momentum around ideas or products -- ranging from kickstarting political campaigns to promoting brands and lifestyle products. A number of occupations, especially in media and politics, increasingly rely on practitioners being successful on social media -- journalists and electoral candidates, for instance, can expect their success at work to be either bootstrapped or bolstered by their Twitter, Instagram, Facebook, YouTube, or even TikTok presences. Social media influencers have also contributed to the overall changes in the contemporary information environment, including the growth of misinformation and polarizing bias. Consequently, the role of influencers is critical in analysing questions about the recent and ongoing changes in media ecologies. 

By building a database of Indian influencers, We make the following main contributions:
\begin{compactitem}
    \item An annotated set of 11580 unique accounts, manually verified as influencers.
    \item Annotations with 7 broad categories and 24 subcategories encompassing a variety of influencers, including an annotation of ``individual" and ``entity" type for each subcategory, usable for various research purposes.
    \item The methods used in this dataset can be used to build similar datasets for other groups of accounts -- arranged by nation, state or domain.
    \item The first major attempt at making an influencer dataset set in the Global South made available for public use.
    \item This is the first large dataset, to our knowledge, that allows aggregated analysis on accounts categorized by industry and occupations, covering a vast majority of influencers in public life within an ecosystem.
    
\end{compactitem}

Our dataset is available on Dataverse (DOI: \href{https://doi.org/10.7910/DVN/BPY2JY}{https://doi.org/10.7910/DVN/BPY2JY})

\section{Related Work}

% Work in Progress
Twitter has been of great interest as a data-source to researchers. The Twitter API provides an easy interface to access user and tweet information. However, critical user attributes, such as geo-location, gender and political leanings are not accurately or reliably represented in the Twitter API. To this end, there have been many attempts at annotating these attributes for users. \citet{Cheng2010} proposes a Machine Learning approach based exclusively on the tweets made by a user, in the absence of all geospatial clues. \citet{Mahmud2021} builds on the prior literature by introducing a hierarchical classifier to improve prediction accuracy. 

Other attributes, such as gender, age, ethnicity and political affiliation have also been of recurring interest in the research community. \citet{Conover_2011} uses an SVM on a manually annotated set of training labels to predict political affiliation of users based on their tweets and hashtag usage. \citet{Pennacchiotti_2021} predicts political orientation and ethnicity by leveraging other observable information such as Twitter networks and user behaviour. \citet{Al_2021} attempts to infer these attributes from the neighbours (friends) of a user - an approach based on similar underlying assumptions as ours.Unfortunately, a recent study by \citet{Cohen_2021} claims that most attempts at annotating political affiliations using Machine Learning systematically report overoptimistic accuracies (nearly 30\% higher than expected) due to the way validation datasets are built. 

\section{Data Collection}
We started the data collection process with the assumption that a seed set can be built using twitter accounts that major politicians choose to follow, using the logic that politicians will follow other politicians or accounts that have some importance in the public sphere - such as media, journalists, key influencers etc. To do this, we used an existing list of known political accounts in India \citet{inproceedings}, culled the list of their friends using the Twitter API, and removed all known politicians from the thus exapnded set. From the remainder, we manually removed non-Indian accounts (i.e. accounts that originated from, or relating to an non-Indian person, defined as someone primarily known for their activity within India).  

This process resulted in  100k+ accounts from which we removed accounts that are not followed by at least three users from the initial seed set, in order to eliminate accounts that were highly likely to be one-off friends of individuals on the seed set. This was verified manually. At the end of this process, we were left with approximately 10k Twitter accounts of potential influencers that are highly followed by Indian politicians.

This initial set was coded manually, thereupon we did repeat iterations to find accounts followed by journalists, using the assumption that journalists follow newsmakers. We found after two iterations that while we were able to cover a majority of public figures with active twitter accounts. We did this through an exercise of various team members seeking out entertainers, sportspersons, journalists to check if they made the list. We found that the process biased our sample towards more Hindi-speaking states, since they dominate the national narrative. 
\newline \indent
To mitigate this and ensure more representation among regional states, we ran the same process with journalists and politicians in regional states to ensure more equitable coverage. This process nonetheless has certain disadvantages -- a sampling process starting with a seed set of sportspersons or businesspersons would for instance  have a relatively larger set of people in sports or business and so on. We used politicians and journalists since they are generally interested in influential individuals across domains.
%While our initial process is built ground up from friends of politicians, it implies that our definition of influencer is biased towards those accounts that are of interest to politicians - i.e. for instance, if our seed set had started with sportspersons, it is likely we would have ended up with several highly followed sports management companies or commentators in our set, or likewise for other domains. 
\newline \indent
The team’s contextual knowledge of India is used to ensure that known public figures are included in the sample as far as possible. Our process therefore leans more heavily towards influencers with a general appeal than those who are very specific to a field. For instance, a journalist with as few as 3000 Twitter followers may be included in our list, while a film PR agent with 100k followers may end up excluded if they did not have followers from across domains. We also apply the primary domain of engagement by an individual at any given point. Thus, if a person is primarily known for their sporting activities, but also has business interests (such as cricket players Virat Kohli or MS Dhoni) they are nonetheless considered sportspersons in our sample. Likewise a journalist of repute who has written a book or has a leadership role in a media entity, such as Rajdeep Sardesai or Paranjoy Guha Thakurta, would nonetheless be classified as a journalist rather than as an author or businessperson.

\section{Applications of the Data}\label{application}
Our goal of making this data public is to allow researchers to use a comprehensive list of influencers to study various forms engagement and influence on social media on topics ranging from brand management and political outreach to dangerous speech and disinformation in the Indian context. This database is uniquely flexible as the the categories and subcategories can be expanded/collapsed based on the use case. The methodology can be iteratively repeated to get a highly curated list of required categories. Some applications of the database in past work have been described below:

\begin{compactitem}
\item \citet{dash2021divided} explore influencer polarisation during political crises in India and find that influencers engage with the controversial topics in a partisan manner, with polarized influencers being rewarded with increased retweeting and following. They also observe that specific groups of influencers, particularly fan accounts and platform celebrities consistently engage in polarizing behavior online, thereby underscoring the importance of influencers in political discourse on social media platforms. 
% mainly how various categories of influencers are key in information dissemination and exacerbating political polarisation in various contexts. They study how certain categories of influencers are consistently polarised across all political events more than the others, and bring about the essential role played by these accounts along with politicians. 

\item In another study that explored the manifestations of extreme speech through a case study of violent protests and policing in the city of Bangalore, provoked by a derogatory Facebook post, \citet{dash2021extremism} found that influential accounts were central in manipulating the discourse surrounding the incident. The dominant narratives that were propagated, employed whataboutism to deflect attention from the triggering post and serve as breeding grounds for religion-based extreme speech.

\item The role of influential accounts in disseminating dangerous speech on social media was studied in \cite{dash2021insights}, where they identified dangerous speech by influential accounts on Twitter in India around three key events, examining both the language and networks of messaging that condoned or actively promoted violence against vulnerable groups. They found that dangerous users are more active on Twitter as compared to other users as well as most influential in the network, and act as “broadcasters” in the network, where they are best positioned to spearhead the rapid dissemination of dangerous speech across the platform. 

\item \citet{mishra2021sports} uses our dataset as a seed set for sportspersons and goes over multiple iterations of the workflow to obtain a list of highly influential sports accounts in India. They apply the same methodology to curate a similar database for sportspersons in the USA and do a comparative analysis of the engagements of sportspersons with politicians between the two countries. Similar methodology was used by \citet{10.1145/3460112.3471983} to curate a dataset of influential business leaders in the USA, and did an analysis of their commentary on key issues related to Sustainable Development Goals on social media compared to influential business leaders in India, curated by snowballing the list in our database.

\item \citet{mothilal2022voting} examine the Twitter engagement between Indian politicians and two sub-categories of influencers in our dataset - 'entertainers' and 'sportspersons'. They propose metrics to measure partisanship along different modes of engagement, analyze the discourse in engaged tweets, and study the public reception of such engagements. They find that the ruling party was more effective in reaching out to celebrities by shunning explicit partisan topics and subtly employing non-partisan narrative technique instead.
\end{compactitem}

\section{Data Description}

The dataset we present has a main table with 11580 records. The table has 11 columns, the details of which are present in Table \ref{table: columns}. Among them, our novel contribution are the fields of ``category", ``sub-category" and ``type". Categories refer to a typology of the individuals based typically on their occupation. Further, to account for the presence of influencers from non-traditional backgrounds of celebrity, we refer to the ``digital first" category. This refers to accounts that share little about their offline lives, and owe their popularity to their digital activity. For instance, the sub-categorisation of ``fan accounts", ``humour" (relating to meme accounts), and ``informational" (e.g. an account on automated weather updates) bring forth the nature of the broad category. 

Here, the ``platform first" sub-category is of particular interest, as it contains numerous accounts whose offline lives are either unverifiable, or never mentioned in the first place. These accounts may often claim to hold certain professions, but almost exclusively chime in on political controversies or news on the platform. As studies of this dataset have indicated \cite{dash2021divided, dash2021insights}, the continued involvement of platform celebrities in furthering partisan viewpoints raise further questions about grey area between layperson commentary and coordinated topic manipulation on Twitter. 

In addition to the ``digital first" category, the remaining broad fields we consider in the dataset have to do with the types of authority ascribed to commentators. Thus, ``media" here refers to those involved in the creation and production of news, while ``creatives" is a category for accounts engaged in arts, literature, film and TV production. The ``civil society" category is distinct from the governmental and business realm and is ideally expected to to contribute to public discussions in an informative manner, relating to their specialist domains. Thus, lawyers, doctors, academics make up the category, apart from special interest groups and religious bodies. The last two categories of ``sport" and ``business" represent the fairly straightforward relationship these accounts have with the two occupations, either as organisations involved in the daily workings of the fields, or as individuals engaged in them. 

Additionally, depending on the nature of studies that utilize this dataset, particular sub-categorisations can also be paired together for the unique requirements of research questions.  For instance, a study on sporting commentary can include both the ``sports" category, as well as journalists who contribute to public discussions about matches. Studies on policing can also refer specifically to the ``law \& policy", and ``social worker" sub-categories to track the interaction of law enforcement accounts and advocacy campaigns online. 

A detailed list of these categorical variables are presented in Table \ref{table: subcategories}. Descriptions for each of these subcategories are presented in Table \ref{table: desc} (see below). 

\begin{figure}[!htb]
  \centering
  \includegraphics[width=\columnwidth]{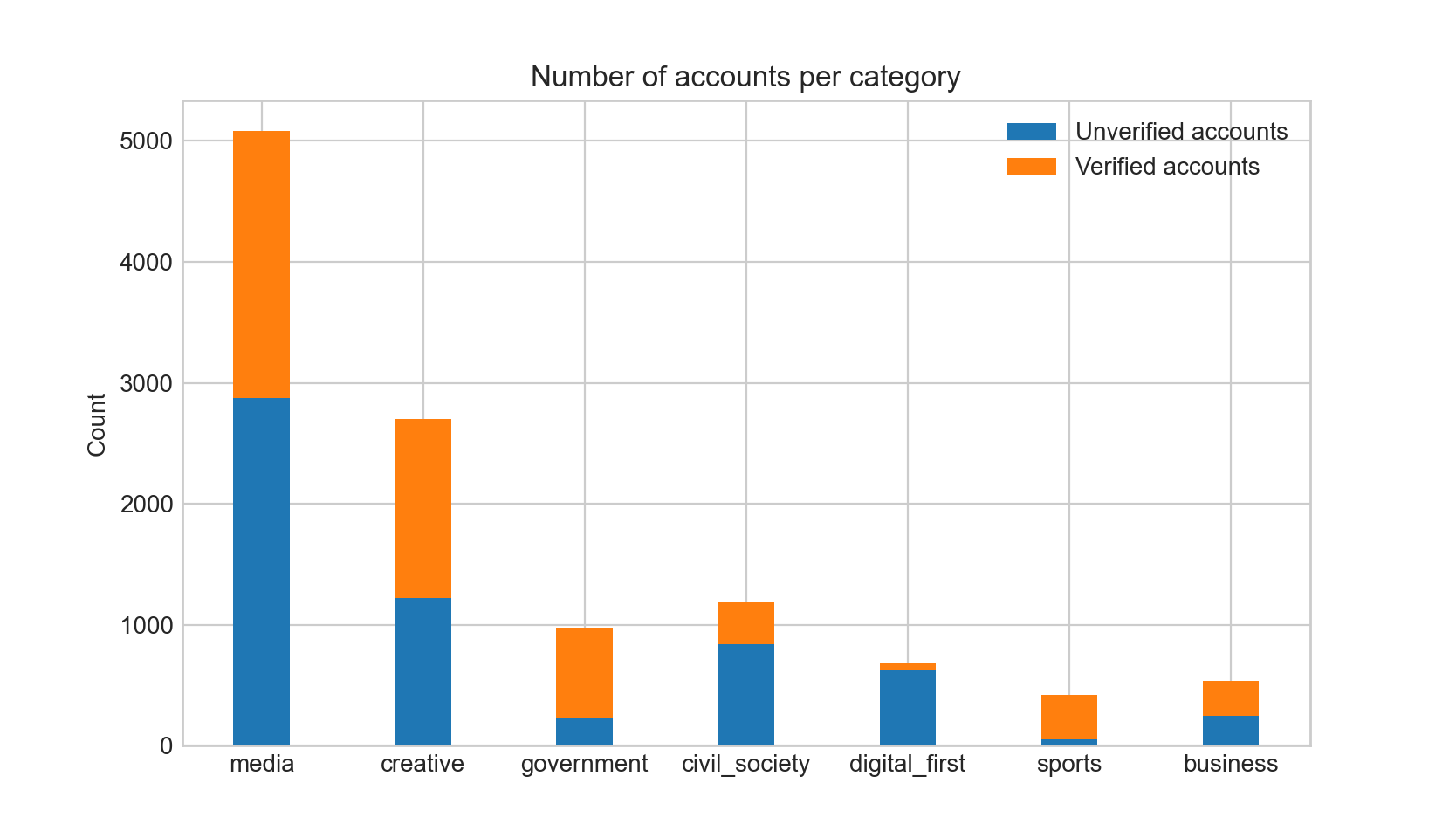}
  \caption{\textbf{Category-Wise Counts}}
  \label{fig: counts}
\end{figure}

\begin{table}[!htb]
\begin{center}
\begin{tabular}{ |c|c|c|c| } 
\hline
\textbf{Category} & \textbf{Counts} & \textbf{Subcategory} & \textbf{Counts} \\
\hline
\multirow{2}{*}{Media} & \multirow{2}{*}{5079} & Journalist & 4099 \\ \cline{3-4}
&& Media House & 980 \\ \hline
\multirow{3}{*}{Creative} & \multirow{3}{*}{2698} & Entertainer & 2622 \\ \cline{3-4}
&& Writer & 66 \\ \cline{3-4}
&& Designer & 10 \\  \hline
\multirow{6}{*}{Civil Society} & \multirow{6}{*}{1184} &  Specialist & 732 \\ \cline{3-4}
&& Law \& Policy & 160 \\ \cline{3-4}
&& Social Worker & 130 \\ \cline{3-4}
&& Research Org & 77 \\ \cline{3-4}
&& Academic & 69 \\ \cline{3-4}
&& Religious Org & 16 \\  \hline
\multirow{5}{*}{Government} &\multirow{4}{*}{977} & Police & 205 \\ \cline{3-4}
&& Bureaucrat & 223 \\ \cline{3-4}
&& Organisation & 405 \\ \cline{3-4}
&& Defence & 29 \\  \cline{3-4}
&& Official & 115 \\ \hline
\multirow{4}{*}{Digital First} & \multirow{4}{*}{684} & Platform First & 431 \\ \cline{3-4}
&& Fan Account & 91 \\\cline{3-4}
&& Informational & 100 \\ \cline{3-4}
&& Humor & 62 \\  \hline
\multirow{2}{*}{Sports} & \multirow{2}{*}{423} & Team & 82 \\ \cline{3-4}
&& Person & 341 \\  \hline
\multirow{2}{*}{Business} & \multirow{2}{*}{535} & Brand & 146 \\ \cline{3-4}
& & Person & 389 \\ \cline{3-4}
\hline
\multirow{1}{*}{\textbf{Total}} & \multirow{1}{*}{11580} &   & 11580 \\  \hline

\end{tabular}
\caption{\textbf{Subcategory-Wise Counts}}\label{table: subcategories}
\end{center}
\end{table}

\begin{table*}[!htb]
\begin{center}
\begin{tabular}{ |c|c|c|c| } 
\hline
\textbf{Field Name} & \textbf{Description} & \textbf{Type} & \textbf{Unique Counts}\\
\hline
id\_str & Unique Twitter ID & Unique String & 11580\\\hline
created\_at & Account creation date & UTC Datetime & 11578\\\hline
name & Twitter name & String & 11442\\\hline
username & Twitter Handle & Unique String & 11580\\ \hline
description & Twitter bio & String & 11124\\\hline
followers & Number of followers & Numeric & 10081 \\\hline
url & URL from profile text & String & 7515\\\hline
location & Location & String/Categorical & 2408\\\hline
type & Individual or Entity & Categorical & 2\\\hline
verified & Blue-ticked account & Boolean & 2\\\hline
category & Primary Industry & Categorical & 7\\\hline
sub\_category & Primary Occupation & Categorical & 24\\\hline
\end{tabular}
\caption{\textbf{Dataset Columns}}\label{table: columns}
\end{center}
\end{table*}

\section{Analysis of Represented Accounts}

%\begin{figure}[h]
%\begin{tabular}{cc}
%  \includegraphics[width=0.2\textwidth]{LaTeX/tables/civil.png} &   %\includegraphics[width=0.2\textwidth]{LaTeX/tables/creative.png} \\
%  (a) Civil Society & (b) Creative \\
%  \includegraphics[width=0.2\textwidth]{LaTeX/tables/govt.png} &   %\includegraphics[width=0.2\textwidth]{LaTeX/tables/digital.png} \\
% (c) Government & (d) Digital First \\[6pt]
% \caption{Category-wise distribution of users}
%\end{tabular}
%\end{figure}

\begin{figure}[!htb]
  \centering  \includegraphics[width=\columnwidth]{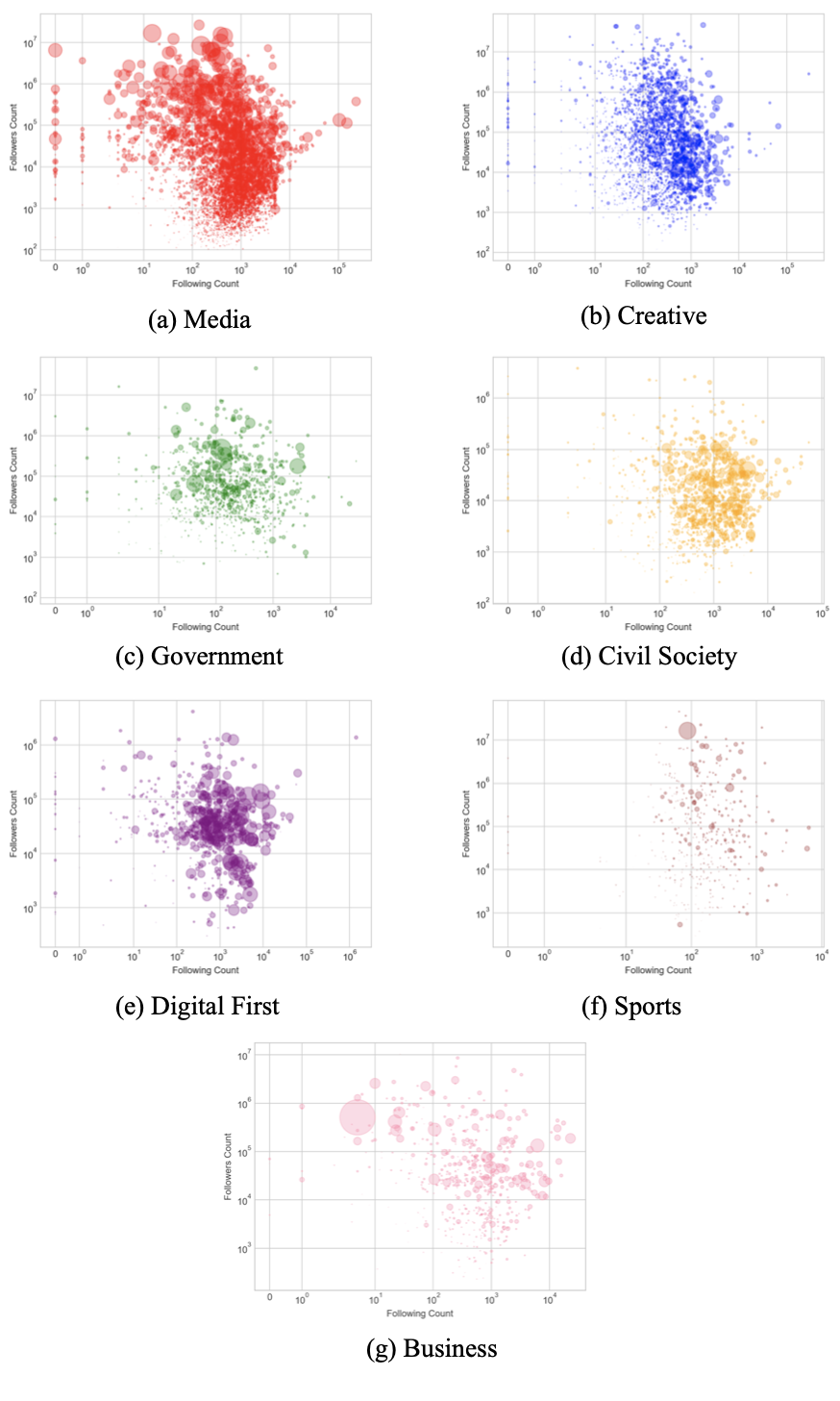}
  \caption{\textbf{Visualising 'measures-of-influence' of represented users across all categories.} The x-axis represents friends, y-axis represents followers and the size of a bubble is proportional to the tweets volume}
  \label{fig: scatter}
\end{figure}

We visualise all the Twitter users represented in out dataset across the 7 categories as separate category-wise scatterplots in Figure \ref{fig: scatter}. To locate a user in a 2-dimensional plane, the x-axis (logarithmic) represents the number of accounts followed by a user and the y-axis (logarithmic) represents the number of followers of the user. The size of the point representing each user is proportional to the total number of tweets and retweets ever made by that user. All these public metrics were accessed via the Twitter API. We observe from the visualizations that:
\begin{compactitem}
    \item Users categorized as Civil Society tend to have narrower range of followers ($10^3 - 10^5$) than those categorized as Media or Creative ($10^3 - 10^7$)
    \item In contrast to the other categories, many Business accounts that typically tweet most actively tend to follow much fewer people ($< 10^2$)
\end{compactitem}

\section{Ethical Considerations}
\subsection{Composition}
The dataset contains 11580 records, each of which are of Twitter users. Additional details of its composition has been described in an earlier section and in Tables \ref{table: subcategories},\ref{table: columns} and \ref{table: desc}. By its nature, the dataset is a non-random sample from the set of all Twitter Users. Only 2 fields ('url' and 'location') may have missing information for some instances - these are self-reported values and their availability depends entirely on whether the Twitter user has shared any such value. All other fields have no missing values. By it's nature, the dataset can be used to identify individuals, more specifically, Twitter users. Apart from certain self-reported fields, such as location - no other field contains any sensitive information. All fields in the dataset are derived from the Twitter API and is public to general audiences.
\newline \indent
Since the dataset involves manual annotation of the ``category" and ``subcategory" fields, a small margin of human-error is to be expected. In the absence of any real measure of ground-truth, we are unable to evaluate this error percentage quantitatively. 

\subsection{Uses}
This dataset, subsets and expansions thereof, have been used, and are currently being used in several academic projects. We list some published research using this dataset in a prior section. We also outline other possible uses in the aforementioned section on applications. We note that all uses of the dataset must be cognizant of unavoidable errors that may have crept in as a result of manual annotation. Certain fields (``username",``description",``followers" etc.) may change over time -- we recommend updating these fields using the Twitter API and the ``id\_str" field before the dataset is used in any application involving these fields. We reiterate that the primary contribution of this dataset is the annotation of ``categories" and ``subcategories".
\subsection{Distribution and Maintenance}
We have hosted the dataset on Dataverse. It may be accessed directly from Dataverse for any uses. Our main fields of interest are 'category' and 'subcategory' - these will not be updated unless to correct labelling errors. In case of the latter, kindly contact the authors. By it's nature, this dataset may be augmented and expanded upon by users to tailor it to their specific needs. The authors cannot guarantee or verify such modifications, however.
\subsection{FAIRness  of the Dataset}
We host our dataset (along with metadata) on Dataverse. Dataverse is an open-source data repository software used widely, which provides a convenient way for dataset authors to adhere to FAIR \cite{FAIR} principles. Our attempt at the same involves:
\begin{compactitem}
    \item \textbf{Findability}: Dataverse assigns a unique DOI (Document Object Identifier) when a dataset is published. This DOI resolves to a landing page with metadata, data files, terms, waivers and licenses.
    \item \textbf{Accessibility}: Dataverse provides public machine-accessible interfaces to search the data, access the metadata and download the data files, using a token to grant access when data files are restricted (‘A’).
    \item \textbf{Interoperability and Resuablility}: Dataverse offers the metadata at following 3 levels of hierarchy: 
    \begin{enumerate}
        \item data citation metadata (DataCite or Dublin Core)
        \item domain-specific metadata
        \item file-level metadata
    \end{enumerate}
\end{compactitem}

\section{Conclusion and Future Work}
A novel database of highly networked individuals on Twitter is a window to better our understanding of narrative and influence on Social Media. With our focus on the Global South, we believe this dataset opens up new possibilities to understand interaction and dialogue in India, especially on how influencers in various spaces of public life intersect with vested interests such as politicians, or impact the public discourse, such as influencing the conversation on certain topics. Along the lines of previous work that have used subsets of this dataset, we are also working on analysing various other categories of users present in our dataset, such as government and defence.

This dataset and its sub-categories are also meant to be a living resource, since new influencers will get added, and categories will not only need to be updated for specific accounts, but the entire notion of a category may need to be rethought, reframed. What we also do here is provide a reasonably exhaustive, and closely vetted collection of  seed-accounts which can be used to iteratively build a larger set by snow-balling through their immediate friend networks. For instance, if one were interested in dramatically increasing the ``business" category, one could quickly snowball that into a much larger set to do a deeper study of business behavior on social media in India. This opens up the possibility for deeper dives into running domain-specific studies that need to characterize how an entire universe of users in a category behave on social media. We plan on using similar techniques to build upon specific sub-categories and study their interactions in greater detail.

\begin{table*}[!htb]
\centering
 
\begin{tabular}{|p{0.15\textwidth}| p{0.2\textwidth}| p{0.5\textwidth}|}
\hline
\textbf{Categories} & \textbf{Sub Categories}     & \parbox[c]{1 cm}{\raggedright \textbf{Description}}  \\ 
\hline
\multirow{6}{*}{civil society}       & academic           & academic/research publications separate from newspapers and magazines, individuals in academia, or individuals engaged in fiction or non-fiction authorship.\\\cline{2-3}
                    & social worker      & individuals who are distinguished by their work and service towards social causes \\ \cline{2-3}
                    & law  policy        & non-academic specialists engaged with policy and law-making  \\ \cline{2-3}
                    & religious organisation   & entities associated with religious and spiritual sects  \\\cline{2-3}
                    & research organisation  &  entities associated with academic research - includes publishing houses and think tanks  \\\cline{2-3}
                    & specialist  &  individuals who have highly specialised areas of work and impact, which aren't covered in the other sub-categories  \\ \hline

\multirow{4}{*}{government}          & defence            &  Accounts of defence personnel - veterans and other military professionals\\ \cline{2-3}
                    & bureaucracy        & Accounts of Indian bureaucrats, IAS, and state-level administrative officers. (E.g.: DMs, Ministry Secretaries etc.) (IFS not included)  \\ \cline{2-3}
                    & police             & District, state, railway, traffic, police accounts, control rooms. \\ \cline{2-3}
                    & official    & Nationalised entities such as national banks, administrative departments, NITI Ayog, branches of defence etc. Also includes official accounts of people with positions in the government \\\hline

\multirow{2}{*}{business}       & brand              & Major product brand, differentiated from a product holding company or corporation. \\ \cline{2-3}
                    & businessperson     & Individual owner or leader of a business concern. \\ \hline
\multirow{2}{*}{media}      & journalist           & news anchors, columnists and others at major media houses \\ \cline{2-3}
                    & media house          & entities and organisations involved in journalism, news, media and advertising \\ \hline

\multirow{3}{*}{creative}  & entertainment      & includes artists, actors, musicians, comedians, reality TV talent, show hosts \\ \cline{2-3}
                    & writer    & authors, primarily deriving fame outside academic circles \\ \cline{2-3}
                    & designer    & artists and designers of prominence \\ \hline
\multirow{2}{*}{sports}              & team        &   official accounts of sports teams - includes cricket, football, IPL teams etc \\\cline{2-3}
                    & person        &    accounts of sports persons\\\hline

 \multirow{4}{*}{digital first}       & fan account        & accounts in appreciation of celebrities \\\cline{2-3}
                    & humour             & satire, parody, meme pages and accounts \\\cline{2-3}
                    & informational      & includes accounts sharing facts, information, historical accounts, job opening updates etc.\\\cline{2-3}
                    & platform first    & influencers for whom the primary source of popularity is through an internet platform (e.g.: YouTubers, TikTokers, Twitter-famous civilians.)\\\hline
\end{tabular}
\caption{\textbf{Subcategory Description}}\label{table: desc}
\end{table*}

%\bibliographystyle{plain}
%\bibliography{reference}

\begin{quote}
\begin{small}
\bibliography{reference}
\end{small}
\end{quote}

\end{document}